\documentclass{iopconfser}
\usepackage{graphicx}
\usepackage{cite}
\begin{document}
\def\ave#1{\langle#1\rangle} 

\title{Magnetization Plateau of the $S={1 \over 2}$ Distorted Diamond Spin Chain with Ferromagnetic Interaction}

\author{Masaru Hashimoto$^{1}$, Koki Doi$^{1}$, Tomoki Houda$^1$, Rito Furuchi$^1$, Hiroki Nakano$^{1}$, Kiyomi Okamoto$^1$, T\^oru Sakai$^{1,2}$}

\affil{$^1$School of Science, University of Hyogo, Hyogo 678-1297, Japan}
\affil{$^2$National Institutes for Quantum Science and Technology, SPring-8, Hyogo 679-5148, Japan}

\email{sakai@spring8.or.jp}

\begin{abstract}
The magnetization process of the $S=1/2$ distorted diamond spin chain with ferromagnetic interactions is investigated 
using the numerical diagonalization of finite-size clusters.
The level spectroscopy analysis applied for the model with the spin anisotropy 
indicates that two different magnetization plateau phases appear at 1/3 of the saturation magnetization. 
The phase diagrams for some typical interaction parameters are presented. 
In addition the magnetization curves for several typical parameters are obtained. 
\end{abstract}

\section{Introduction}
The distorted diamond spin chain \cite{okamoto1999,honecker,okamoto2003,kikuchi,kikuchi2,gu-su,honecker2,jeschke,
ananikian,ueno,filho} 
is one of interesting low-dimensional magnets. 
It is known as a theoretical model of the material Cu$_3$(CO$_3$)$_2$(OH)$_2$, called azurite\cite{kikuchi}. 
One of characteristic features of this compound is the magnetization plateau at 1/3 of the saturation magnetization, 
which was theoretically predicted \cite{honecker,okamoto2003} prior to the experimental finding\cite{kikuchi,kikuchi2}. 
Recently another candidate compound for the distorted diamond spin chain K$_3$Cu$_3$AlO$_2$(SO$_4$)$_4$, 
called alumoklyuchevskite, was discovered\cite{fujihara,fujihala2017}. 
Although this material includes some ferromagnetic interactions, different from azurite, 
it is also theoretically predicted to exhibit the 1/3 magnetization plateau\cite{morita}. 
Then it would be interesting to consider the theoretical mechanism of the 1/3 magnetization plateau 
in the $S=1/2$ distorted diamond spin chain with some ferromagnetic interactions. 
In the previous works on this model including a ferromagnetic interaction per unit cell, it was revealed 
to exhibit the 1/3 magnetization plateau due to the ferrimagnetic mechanism\cite{sakai1}, 
as well as the field-induced spin nematic liquid phase\cite{sakai2}.
In this work, we investigate the magnetization process of this model including two ferromagnetic interactions per unit cell, 
using the numerical exact diagonalization of finite-size clusters and the level spectroscopy analysis\cite{kitazawa, nomura-kitazawa,okamoto2}. 
As a result it is found that when some interaction parameters are varied, two different magnetization plateau phases
appear at 1/3 of the saturation magnetization. 
One of the two phases corresponds to the symmetry protected topological phase\cite{gu,pollmann}.
The phase diagram including these two plateau phases, as well as the no-plateau phase, at the 1/3 magnetization is presented.

\section{Model and calculations}

\begin{figure}[h]
\centerline{\includegraphics[width=0.40\linewidth,angle=0]{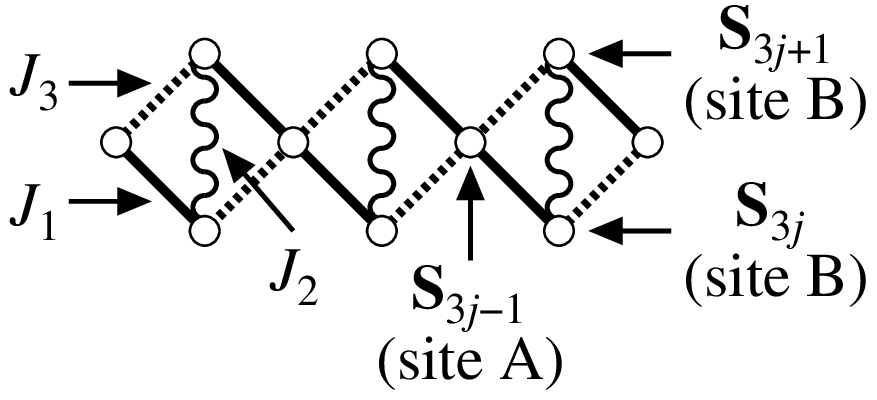}}%
\caption{\label{model}
$S=1/2$ distorted diamond chain. $J_1$ is ferromagnetic and $J_2$ and $J_3$ are antiferromagnetic. 
There are two kinds of sites, site A and site B.}
\end{figure}

We consider the magnetization process of the $S=1/2$ distorted diamond chain as shown in Fig. \ref{model}. 
We fix $J_1$ to $-1$ (ferromagnetic) and consider the case of the antiferromagnetic $J_2$ and $J_3$ bonds. 
The $XXZ$ anisotropy is also introduced to the $J_3$ bond. 
The Hamiltonian has the form
\begin{eqnarray}
\label{ham}
{\cal H}&=&{\cal H}_0+{\cal H}_Z, \\
\nonumber
{\cal H}_0& =& J_{\rm 1} \sum_{j=1}^L(\vec{S}_{3j-1}\cdot\vec{S}_{3j} + \vec{S}_{3j}\cdot\vec{S}_{3j+1})  + J_2 \sum_{j=1}^L\vec{S}_{3j+1}\cdot\vec{S}_{3j+2} \\
 &&+J_3\sum_{i=1}^L \left[(S_{3j-2}^xS_{3j}^x + S_{3j-2}^yS_{3j}^y) + \lambda  S_{3j-2}^zS_{3j}^z
 + (S_{3j}^xS_{3j+2}^x + S_{3j}^yS_{3j+2}^y) + \lambda S_{3j}^zS_{3j+2}^z \right],  \\
{\cal H}_Z &=&-H\sum_{j=1}^{3L} S_{j}^z,
\end{eqnarray}
where $S_i^\mu$ denotes the $\mu$ component of the $S=1/2$ operator at the $i$th site,
$H$ is the external magnetic field along the $z$-direction, 
and $L$ is the number of unit cells (such that there are $N=3L$ spins in the system). 
Using the Lanczos method, we calculate the lowest energy $E(L,M)$ in the subspace where $\sum _j S_{j}^z=M$ 
for each $L$, under the periodic boundary condition. 
The reduced magnetization $m$ is defined by $m=M/M_{\rm s}$, 
where $M_{\rm s}$ denotes the saturation of the magnetization, 
namely $M_{\rm s}=3L/2$. 

\section{1/3 magnetization plateau}

\begin{figure}[h]
   \begin{minipage}{.48\linewidth}
      \begin{center}
         \scalebox{0.4}[0.4]{\includegraphics{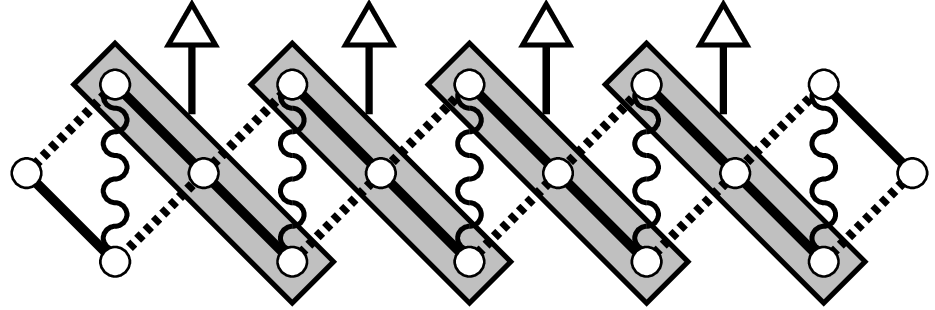}}
      \end{center}
\caption{\label{3-ferro}
Schematic picture of the ferromagnetic 3-spin cluster mechanism of the 1/3 magnetization plateau. 
Grey rectangles denote ferromagnetically coupled 3-spin cluster with the total of $S^z$ is equal to $1/2$.
It is realized in the limit $|J_1| \gg J_2, J_3$. 
}
   \end{minipage}
   \hspace{0.04\columnwidth}
   \begin{minipage}{.48\linewidth}
      \begin{center}
         \scalebox{0.4}[0.4]{\includegraphics{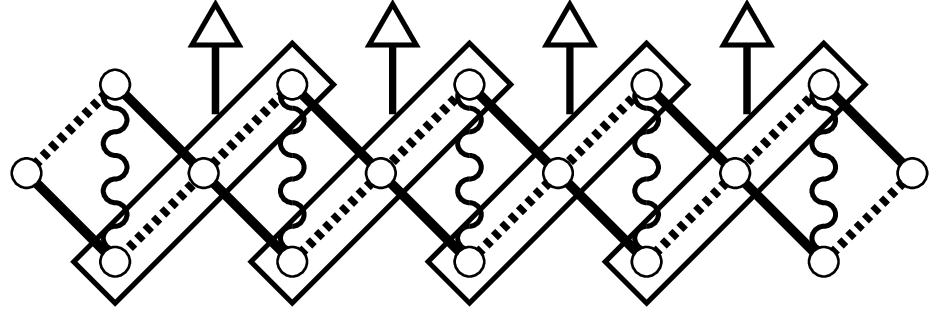}}
      \end{center}
\caption{\label{3-af}
Schematic picture of the antiferromagnetic 3-spin cluster mechanism of the 1/3 magnetization plateau. 
White rectangles denote antiferromagnetically coupled 3-spin cluster with the total of $S^z$ being equal to $1/2$.
It is realized in the limit  $J_3 \gg |J_1|, J_2$. 
}
   \end{minipage}
\end{figure}

\begin{figure}[h]
      \begin{center}
         \scalebox{0.4}[0.4]{\includegraphics{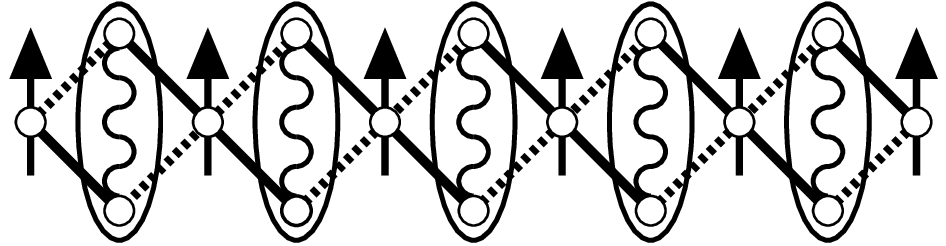}}
      \end{center}
\caption{
\label{dimer}
Schematic picture of the  dimer + free spin mechanism of the 1/3 magnetization plateau. 
Ellipses denote singlet dimer pairs of spins.
It is realized in the limit $J_2 \gg |J_1|, J_2$. 
}
\end{figure}

\begin{figure}[h]
\bigskip
   \begin{minipage}{.48\linewidth}
      \begin{center}
         \scalebox{0.32}[0.32]{\includegraphics{m1-3LSisoj201.eps}}
      \end{center}
\caption{\label{level}
Three gaps, $\Delta_2$ (dashed curves), $\Delta_{\rm TBC+}$ (dash-dot curves)  and $\Delta_{\rm TBC-}$ (solid curves) 
are plotted versus $J_3$ for $L=4$ (black), 6 (red) and 8 (green) in case of $J_2 = 0.01$ and $\lambda = 1.0$. 
}
   \end{minipage}
   \hspace{0.04\columnwidth}
   \begin{minipage}{.48\linewidth}
      \begin{center}
         \scalebox{0.32}[0.32]{\includegraphics{gaisodm12j201.eps}}
      \end{center}
\caption{\label{gaiso}
$L \to \infty$ extrapolation of the crossing point between $\Delta _2$ and $\Delta_{\rm TBC-}$ for $L=4$, 6 and 8.
We estimate the 
critical point $J_{3c}$ in the thermodynamic limit as $J_{3c}=0.192$. }
   \end{minipage}
\end{figure}

\begin{figure}[h]
\bigskip
\medskip
\centerline{\includegraphics[width=0.50\linewidth,angle=0]{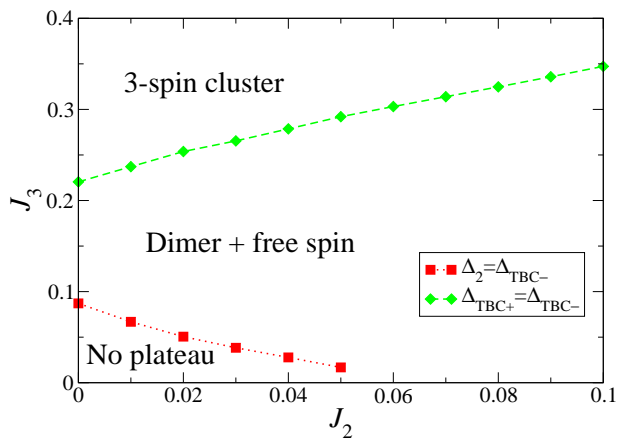}}%
\caption{\label{phase3}
Phase diagram at $m=1/3$ in the case of the easy-axis anisotropy $\lambda =1.2$. 
}
\end{figure}

We focus on the magnetization plateau at $m=1/3$. 
We consider three different mechanisms of the 1/3 plateau as follows: 
(a) the  ferromagnetic 3-spin cluster, (b) the antiferromagnetic 3-spin cluster, 
and (c) the dimer + free spin mechanisms. 
The schematic pictures of these mechanisms are shown in 
Figs. \ref{3-ferro}, \ref{3-af} and \ref{dimer}. 
They are realized in the limits $|J_1| \gg J_2, J_3$, and $J_3 \gg |J_1|, J_2$, and 
$J_2 \gg |J_1|, J_2$, respectively. 
Since the two mechanisms (a) and (b) cannot be distinguished by any symmetry, 
no clear phase boundary would exist between them. 
Thus we take the plateau phase based on these two mechanisms for a single phase 
and call it the 3-spin cluster plateau phase. 
This phase is the so-called trivial phase.
On the other hand, another plateau phase (c),
the dimer + free spin plateau phase,
can be distinguished from the 3-spin cluster plateau phase
by the space inversion symmetry under the twisted boundary condition (see later).
This phase belongs to the symmetry-protected topological phase.
In order to obtain the phase diagram including these plateau and no-plateau phases, 
we use the level spectroscopy method \cite{kitazawa, nomura-kitazawa,okamoto2}. 
According to this analysis, 
we should compare the following three energy gaps; 
\begin{eqnarray}
\label{delta2}
&&\Delta _2 ={E(L,M-2)+E(L,M+2)-2E(L,M) \over 2}, \\
\label{tbc+}
&&\Delta_{\rm TBC+}=E_{\rm TBC +}(L,M)-E(L,M), \\
\label{tbc-}
&&\Delta_{\rm TBC-}=E_{\rm TBC -}(L,M)-E(L,M),
\end{eqnarray}
where $E_{\rm TBC+}(L,M)$ ($E_{\rm TBC-}(L,M)$) is the energy of the 
lowest state with the even parity (odd parity) with respect to the space inversion 
at the twisted bond under the twisted boundary condition applied for the $J_2$ and $J_3$ bonds, 
and other energies are under the periodic 
boundary condition. 
The level spectroscopy method indicates that the smallest gap 
among these three gaps for $M=L=M_{\rm s}/3$ determines the phase 
at $m=1/3$. 
Namely, $\Delta_2$, $\Delta_{\rm TBC+}$ and $\Delta _{\rm TBC-}$ 
correspond to the no-plateau, 3-spin cluster-plateau and dimer+free spin plateau phases, 
respectively. 
In the case of $J_2=0.01$ and $\lambda=1.0$, 
the gaps $\Delta_2$ (dashed curves), $\Delta_{\rm TBC+}$ (dash-dot curves)  and $\Delta_{\rm TBC-}$ (solid curves) 
are plotted versus $J_3$ for $L=4$ (black), 6 (red) and 8 (green) in Fig. \ref{level}. 
It indicates that when $J_3$ increases, the phase changes from the no-plateau to the dimer+free-spin and 
finally to the 3-spin-cluster phases. 
The crossing points between $\Delta _2$ and $\Delta_{\rm TBC-}$ for $L=4$, 6 and 8 are plotted 
versus $1/L$ in Fig. \ref{gaiso}. 
Assuming that the size correction is the function $C_1/L + C_2/L^2$, we estimate the 
critical point $J_{3c}$ in the thermodynamic limit as $J_{3c}=0.192$. 
The extrapolation error cannot be estimated, because only three system sizes are available ($L=4$, 6 and 8).  
Using this method, the phase diagram in the isotropic case ($\lambda =1.0$) is obtained as shown in 
Fig. \ref{phase1}. 
The boundary between the no-plateau and the dimer+free spin plateau phases looks very close to 
that between the two plateau phases on the line $J_2=0$. 
But whether they coincide with each other or not, is not clear, 
because the finite-size effect would be still large for $L=4$, 6 and 8. 
The check by the larger cluster calculation would be necessary to obtain the definite conclusion.
We also obtain the phase diagrams for $\lambda=0.8$ (easy-plane anisotropy) and $\lambda=1.2$ (easy-axis anisotropy), 
as shown in Figs. \ref{phase2} and \ref{phase3}, respectively. 
The tricritical point appears for $\lambda=0.8$.

\begin{figure}[h]
\bigskip
   \begin{minipage}{.48\linewidth}
      \begin{center}
         \scalebox{0.32}[0.32]{\includegraphics{m1-3phase-a310s12.eps}}
      \end{center}
\caption{\label{phase1}
Phase diagram at $m=1/3$ in the isotropic case $\lambda =1.0$. 
}
   \end{minipage}
   \hspace{0.04\columnwidth}
   \begin{minipage}{.48\linewidth}
      \begin{center}
         \scalebox{0.32}[0.32]{\includegraphics{m1-3phase-a308s12c.eps}}
      \end{center}
\caption{\label{phase2}
Phase diagram at $m=1/3$ in the case of the easy-plane anisotropy $\lambda =0.8$. 
}
   \end{minipage}
\end{figure}

%

\section{Site magnetizations}

We also numerically calculate the site magnetizations,
$\ave{S_{\rm A}^z}$ and $\ave{S_{\rm B}^z}$,
for the $\lambda = 1.0$ case
to check the consistency between the physical pictures of plateaus and the phase diagrams of
the previous section.
In case of $|J_1| \gg J_2,J_3$ (mechanism (a) of \S3, Fig. \ref{3-ferro}),
the site magnetizations are expected to
\begin{equation}
     \ave{S_{\rm A}} = \ave{S_{\rm B}} = {1 \over 6},
     \label{eq:sitemag-a}
\end{equation}
whereas for $J_3 \gg |J_1|,J_2$ (mechanism (b) of \S3, Fig. \ref{3-af}),
\begin{equation}
     \ave{S_{\rm A}} = -{1 \over 6},~~~\ave{S_{\rm B}} = {1 \over 3},
     \label{eq:sitemag-b}
\end{equation}
and for $J_2 \gg |J_1|,J_3$ (mechanism (c) of \S3, Fig. \ref{dimer}),
\begin{equation}
     \ave{S_{\rm A}} = {1 \over 2},~~~\ave{S_{\rm B}} = 0.
     \label{eq:sitemag-c}
\end{equation}
The calculated site magnetizations are shown in Figs. \ref{sitemag1} and \ref{sitemag2}.
Their behaviors are consistent with the above predictions.

\begin{figure}[h]
\bigskip
   \begin{minipage}{.48\linewidth}
      \begin{center}
         \scalebox{0.32}[0.32]{\includegraphics{sitemagj201c.eps}}
      \end{center}
\caption{\label{sitemag1}
Site magnetizations at $m=1/3$ for $\lambda =1.0$ and $J_2 = 0.1$. 
The site magnetizations coincide with Eq. (\ref{eq:sitemag-a}) for $J_3 \to 0$
and with Eq. (\ref{eq:sitemag-b}) for $J_3 \to \infty$.
}
   \end{minipage}
   \hspace{0.04\columnwidth}
   \begin{minipage}{.48\linewidth}
      \begin{center}
         \scalebox{0.32}[0.32]{\includegraphics{sitemagj301c.eps}}
      \end{center}
\caption{\label{sitemag2}
Site magnetizations at $m=1/3$ for $\lambda =1.0$ and $J_3 = 0.1$. 
The site magnetizations coincide with Eq. (\ref{eq:sitemag-a}) for $J_2 \to 0$
and with Eq. (\ref{eq:sitemag-c}) for $J_2 \to \infty$.
}
   \end{minipage}
\end{figure}



 \begin{figure}[h]
\centerline{\includegraphics[width=0.50\linewidth,angle=0]{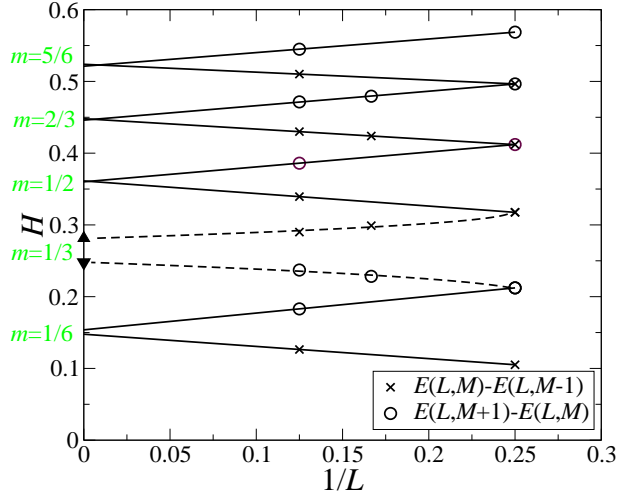}}%
\caption{\label{extra}
$E(L,M)-E(L,M-1)$ (crosses) and $E(L,M+1)-E(L,M)$ (circles) are plotted versus $1/L$ 
for $L=$4, 6 and 8 in the case of $\lambda=1.0$, $J_2=0.03$ and $J_3=0.4$. 
It suggests that these quantities would converge to $H(m)$ in the infinite $L$ limit 
at the gapless $m$. The plateau edges (up and down triangles) at $m=1/3$ 
were estimated by the Shanks transformation. 
}
\end{figure}

\section{Magnetization curves}

It is useful to obtain the theoretical magnetization curve for 
several typical parameters. 
The magnetic fields $H_-(m)$ and $H_+(m)$ are defined as 
\begin{eqnarray}
E(L,M)-E(L,M-1) \rightarrow H_-(m) \quad (L\rightarrow \infty), \\
E(L,M+1)-E(L,M) \rightarrow H_+(m) \quad (L\rightarrow \infty) , 
\label{mag}
\end{eqnarray}
where the size $L$ is varied with fixed $m=M/M_s$. 
If the system is gapless at $m$, the size correction should be 
proportional to $1/L$ and $H_-(m)$ coincides to $H_+(m)$. 
It is justified by Fig. \ref{extra}, where $E(L,M)-E(L,M-1)$ and 
$E(L,M+1)-E(L,M)$ are plotted versus $1/L$ for $\lambda=1.0$, $J_2=0.03$ and $J_3=0.4$. 
It suggests that the system is gapless except for $m=1/3$. 
Then we can estimate $H(m)=H_-(m)=H_+(m)$ using the form
\begin{eqnarray}
E(L,M+1)-E(L,M-1)  \rightarrow H(m) + O(1/L^2).
\label{field}
\end{eqnarray}
If the system has a gap at $m$, the magnetization plateau exists 
and $H_-(m) \not= H_+(m)$. 
In such a case, we use the Shanks transformation\cite{shanks}
\begin{eqnarray}
P'_{L}={{P_{L-2}P_{L+2}-P_L^2}\over{P_{L-2}+P_{L+2}-2P_L}},
\label{shanks}
\end{eqnarray}
where $P_L$ is defined as $P_L\equiv E(L,M)-E(L,M-1)$ and $P_L\equiv E(L,M+1)-E(L,M)$ 
 to estimate  $H_-(m)$ and $H_+(m)$, respectively. 
 This transformation is applied for $L=6$ to estimate the 
 plateau edges $H_-(1/3)$ and $H_+(1/3)$.

\begin{figure}[t]
   \begin{minipage}{.48\linewidth}
      \begin{center}
         \scalebox{0.32}[0.32]{\includegraphics{magj203a10.eps}}
      \end{center}
\caption{\label{curve1}
Magnetization curves for $\lambda=1.0$ and $J_2=0.03$. 
Black circles ($J_3=0.0$), red squares ($J_3=0.2$) and green triangles ($J_3=0.4$) 
 are estimated points in the infinite $L$ limit and lines are guides for the eye. 
}
   \end{minipage}
   \hspace{0.04\columnwidth}
   \begin{minipage}{.48\linewidth}
      \begin{center}
         \scalebox{0.32}[0.32]{\includegraphics{magj206a08.eps}}
      \end{center}
\caption{\label{curve2}
Magnetization curves for $\lambda=0.8$ and $J_2=0.06$. 
Black circles ($J_3=0.0$), red squares ($J_3=0.2$) and green triangles ($J_3=0.4$) 
 are estimated points in the infinite $L$ limit and lines are guides for the eye. 
}
   \end{minipage}
\end{figure}



\begin{figure}[h]
      \begin{center}
         \scalebox{0.32}[0.32]{\includegraphics{magj203a12.eps}}
      \end{center}
\caption{\label{curve3}
Magnetization curves for $\lambda=1.2$ and $J_2=0.03$. 
Black circles ($J_3=0.0$), red squares ($J_3=0.2$) and green triangles ($J_3=0.4$) 
 are estimated points in the infinite $L$ limit and lines are guides for the eye. 
}
\end{figure}

Using these methods, the magnetization curves in the infinite $L$ limit 
for $\lambda=1.0$ and $J_2=0.03$ are obtained in Fig. \ref{curve1}, 
where black circles ($J_3=0.0$), red squares ($J_3=0.2$) and green triangles ($J_3=0.4$) 
are estimated points in the infinite $L$ limit and lines are guides for the eye. 
The 1/3 magnetization plateau appears for $J_3=0.2$ and 0.4. 
We also obtained the magnetization curves for 
$\lambda=0.8$ and $J_2=0.06$ shown in Fig. \ref{curve2}, 
as well as for $\lambda=1.2$ and $J_2=0.03$ shown in Fig. \ref{curve3}. 
We see that the plateaus of red curves are based on the dimer + free spin mechanism
while those of green curves on the 3-spin cluster mechanism
by referring the phase diagrams Figs. \ref{phase1}, \ref{phase2} and \ref{phase3}.

\section{Summary}
The magnetization process of the $S=1/2$ anisotropic distorted diamond chain with the ferromagnetic interaction 
is investigated using the numerical diagonalization of finite-size clusters. 
When the ferromagnetic interaction $J_1$ is fixed to $-1$, 
depending on the parameters of the antiferromagnetic interactions $J_2$ and $J_3$, 
two different magnetization plateau phases are revealed to appear at $m=1/3$. 
The phase diagrams with respect to $J_2$ and $J_3$ are obtained for $\lambda=$1.0, 0.8 and 1.2. 
In addition site magnetizations and several magnetization curves are presented. 
We hope such a magnetization plateau would be observed in some real materials.

\section*{Acknowledgments}
This work was partly supported by JSPS KAKENHI, 
Grant Numbers JP16K05419, JP20K03866, JP16H01080 (J-Physics), 
JP18H04330 (J-Physics), JP20H05274 and 23K11125.
A part of the computations was performed using
facilities of the Supercomputer Center,
Institute for Solid State Physics, University of Tokyo,
and the Computer Room, Yukawa Institute for Theoretical Physics,
Kyoto University.
We used the computational resources of the supercomputer 
Fugaku provided by the RIKEN through the HPCI System 
Research projects (Project ID: hp200173, hp210068, hp210127, 
hp210201, hp220043, hp230114, hp230532, and hp230537).


\end{document}